\documentclass[12pt,preprint]{aastex}

\usepackage{graphicx}
\usepackage{txfonts}
\newcommand{\va}{v_{\mathrm{A}}}
\newcommand{\cs}{c_{s}}

\newcommand{\vap}{v_{\mathrm{A}i}}
\newcommand{\csp}{c_{\mathrm{s}i}}
\newcommand{\ctp}{c_{\mathrm{T}i}}

\newcommand{\vac}{v_{\mathrm{A}e}}
\newcommand{\csco}{c_{\mathrm{s}e}}
\newcommand{\ctc}{c_{\mathrm{T}e}}

\newcommand{\pd}{\partial}

\begin{document}

	\title{Damping of filament thread oscillations: effect of the slow continuum}

	\shorttitle{Damping of filament thread oscillations: effect of the slow continuum}

   \author{R. Soler$^1$, R. Oliver$^1$, J. L. Ballester$^1$, and M. Goossens$^2$}
   \affil{$^1$ Departament de F\'isica, Universitat de les Illes Balears,
              E-07122, Palma de Mallorca, Spain}
              \email{[roberto.soler;ramon.oliver;joseluis.ballester]@uib.es}

  \affil{$^2$ Centre for Plasma Astrophysics and Leuven Mathematical Modeling and Computational Science Center, K. U. Leuven, Celestijnenlaan 200B, 3001 Heverlee, Belgium}
              \email{marcel.goossens@wis.kuleuven.ac.be}

  \begin{abstract}

Transverse oscillations of small amplitude are commonly seen in high-resolution observations of filament threads, i.e. the fine-structures of solar filaments/prominences, and are typically damped in a few periods.  Kink wave modes supported by the thread body offer a consistent explanation of these observed oscillations. Among the proposed mechanisms to explain the kink mode damping, resonant absorption in the Alfv\'en continuum seems to be the most efficient as it produces damping times of about 3 periods. However, for a nonzero-$\beta$ plasma and typical prominence conditions, the kink mode is also resonantly coupled to slow (or cusp) continuum modes, which could further reduce the damping time. In this Letter, we explore for the first time both analytically and numerically the effect of the slow continuum on the damping of transverse thread oscillations. The thread model is composed of a homogeneous and straight cylindrical plasma, an inhomogeneous transitional layer, and the homogeneous coronal plasma. We find that the damping of the kink mode due to the slow resonance is much less efficient than that due to the Alfv\'en resonance. 

  \end{abstract}



\section{Introduction}

The existence of the fine-structure of solar filaments/prominences is clearly shown in high-resolution observations by both ground-based and satellite-on-board telescopes. These fine-structures, here called threads, have a typical width, $W$, and length, $L$, in the ranges $0.2\, {\rm arcsec} < W <  0.6\, {\rm arcsec}$ and $5\, {\rm arcsec} < L <  20\, {\rm arcsec}$ \citep{lin2005}. Filament and prominence threads are believed to be thin magnetic flux tubes partially filled with prominence-like plasma and whose footpoints are anchored in the photosphere. Observers usually report the presence of transverse oscillations and waves in these fine-structures \citep[e.g.][]{lin2003,lin2005,lin2007,okamoto}, which have been interpreted in terms of magnetohydrodynamic (MHD) waves \citep[e.g.,][]{hinode}. The reader is referred to \citet{oliverballester02}, \citet{ballester}, and \citet{banerjee} for reviews of these events. In addition, a typical property of small-amplitude prominence oscillations observed in Doppler series is that they are quickly damped, with damping times of the order of several periods \citep{molowny, terradasobs}. Since small-amplitude oscillations are of local nature and related to periodic motions of threads, there must exist some mechanism which is able to damp thread oscillations in an efficient way. 

The damping of prominence oscillations has recently been investigated in a number of papers \citep[this topic has been reviewed by][]{oliver}. \citet{solerapj} modeled a filament thread as a homogeneous cylinder embedded in an unbounded corona and studied the effect of nonadiabatic mechanisms (thermal conduction, radiative losses, and heating) on the wave damping. As in previous works \citep{carbonell, soler1}, they concluded that thermal effects are only efficient in damping the slow mode, the fast kink mode being almost undamped. Subsequently, \citet{arregui} used a more realistic thread model and included an inhomogeneous transition region between the cylindrical thread and the external medium. He neglected plasma pressure and adopted the so-called $\beta = 0$ approximation, where $\beta$ is the ratio of the plasma pressure to the magnetic pressure. These authors found that the kink mode is resonantly coupled to Alfv\'en continuum modes due to the presence of the transverse inhomogeneous layer, and obtained a ratio of the damping time to the period $\tau_{\rm D} / P \approx 3$ for typical prominence conditions. In the case of coronal loops \citep[see, for example, recent reviews by][]{goossens,goossens08}, resonant absorption has been extensively investigated as a damping mechanism for the kink mode.

Although the plasma $\beta$ in solar prominences is probably small, it is definitely nonzero. Hence, if gas pressure is taken into account in the model of \citet{arregui}, i.e. the $\beta \neq 0$ case, it turns out that the kink mode phase speed, namely $c_k$, is also within the slow (or cusp) continuum that extends between the internal, $\csp$, and external, $\csco$, sound speeds, i.e. $\csp < c_k < \csco$. In this case, the kink mode is not only resonantly coupled to Alfv\'en continuum modes but also to slow continuum modes. The frequency of the kink mode is both within the Alfv\'en and slow continua because of the high density and low temperature of the prominence material in comparison with the coronal values. Hence, the slow continuum damping arises as an additional mechanism to damp the kink mode in filament threads, and its efficiency needs to be assessed. Although the slow resonance has been previously investigated in the context of absorption of driven MHD waves in the solar atmosphere \citep[e.g.,][]{cadez,erdelyi} and sunspots \citep[e.g.,][]{keppens}, the present work studies for the first time the effect of the slow resonance on the damping of the kink mode in filament threads. For coronal loops the kink speed is outside the slow continuum. A coronal loop is presumably hotter and denser than its surrounding corona, so the ordering of sound, Alfv\'en, and kink speeds is $\csco < \csp < \vap < c_k < \vac$. Therefore, there is no slow resonance for the kink mode in coronal loops and the present study does not apply to coronal loops.

This Letter is organized as follows: \S~\ref{sec:math} contains a description of the model configuration and the mathematical method. The results are presented and discussed in \S~\ref{sec:results}. Finally, our conclusions are given in \S~\ref{sec:conclusion}.

\section{Model and method}
\label{sec:math}

\begin{figure}[!htb]
\centering
\epsscale{0.75}
\plotone{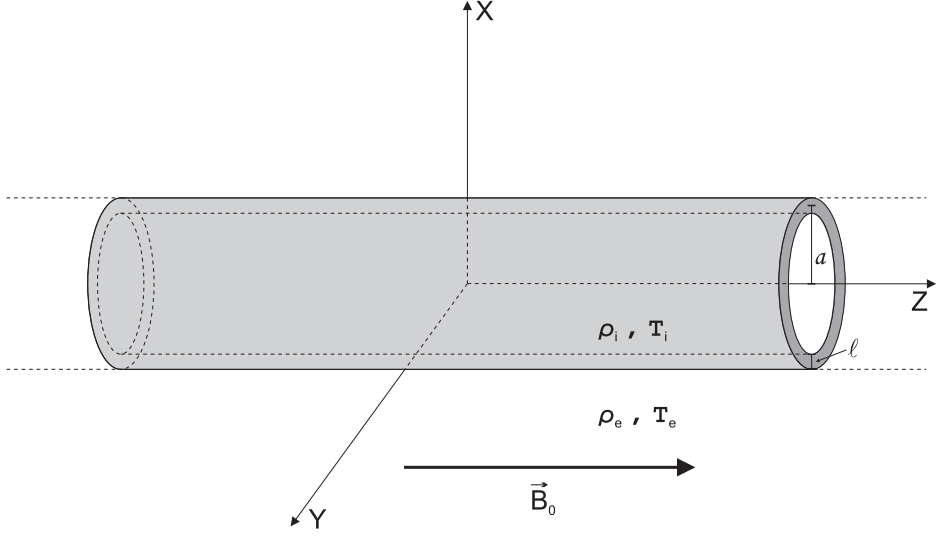}
\caption{Sketch of the model configuration. \label{fig:model}}
\end{figure}

The model for the equilibrium configuration considered here is equivalent to that of \citet{arregui}. We model a filament thread as a straight cylinder with prominence-like conditions embedded in an unbounded corona, with a transverse inhomogeneous layer between both media (see Fig.~\ref{fig:model}). We use cylindrical coordinates, namely $r$, $\varphi$, and $z$ for the radial, azimuthal, and longitudinal coordinates. We adopt the density profile by \citet{rudermanroberts} which only depends on the radial direction,
\begin{equation}
 \rho_{0}\left(r\right)=\left\{\begin{array}{clc}
 \rho_i,&{\rm if}&r\le a - l/2,   \\
 \rho_{\rm tr}\left(r\right),&{\rm if}&a-l/2<r<a+l/2,\\ 
 \rho_e,&{\rm if}&r\geq  a+l/2, \\
\end{array} \right. \label{eq:profile}
\end{equation}
with
\begin{equation}
 \rho_{\rm tr}\left(r\right)=\frac{\rho_i}{2}\left\{\left(1+\frac{\rho_e}{\rho_i}\right) - \left( 1-\frac{\rho_e}{\rho_i}\right)\sin \left[\frac{\pi}{l}\left( r-a\right)\right]\right\}.
\end{equation}
In these expressions, $\rho_i$ is the internal density, $\rho_e$ is the external (coronal) density, $a$ is the tube mean radius, and $l$ is the transitional layer width. The limits $l/a=0$ and $l/a=2$ correspond to a homogeneous tube and a fully inhomogeneous tube, respectively. We use the following densities: $\rho_i=5\times10^{-11}~{\rm kg}~{\rm m}^{-3}$ and $\rho_e=2.5\times 10^{-13}~{\rm kg}~{\rm m}^{-3}$. Therefore, the density contrast between the internal and external plasma is $\rho_i/\rho_e=200$. The plasma temperature is related to the density through the usual ideal gas equation. We consider $T_i=8000$~K and $T_e=10^{6}$~K for the internal and external temperatures, respectively. The magnetic field is taken homogeneous and orientated along the $z$-direction, ${\mathbf B}_0=B_0\hat{\mathbf z}$, with $B_0=5$~G everywhere. With these conditions, $\beta\approx0.04$. 

\subsection{Analytical dispersion relation}

We consider the linear, ideal MHD equations for the $\beta \neq 0$ case. Since $\varphi$ and $z$ are ignorable coordinates, the perturbed quantities are put proportional to $\exp \left( i  \omega t + i m \varphi - i k_z z \right)$, where $\omega$ is the frequency, and $m$ and $k_z$ are the azimuthal and longitudinal wavenumbers, respectively. In the absence of a transitional layer, i.e. $l/a = 0$, the dispersion relation of magnetoacoustic waves is obtained by imposing the continuity of the radial displacement, $\xi_r$, and the total pressure perturbation, $p_{\rm T}$, at $r=a$. This dispersion relation \citep{edwinroberts} is $\mathcal{D}_m  \left( \omega , k_z \right)= 0$, with
\begin{equation}
 \mathcal{D}_m \left( \omega , k_z \right) = \rho_e  \frac{m_i}{\left( \omega^2 - k_z^2 \vap^2 \right)} \frac{J'_m \left( m_i a \right)}{J_m \left( m_i a \right)} 
 - \rho_i  \frac{m_e}{\left( \omega^2 - k_z^2 \vac^2 \right)} \frac{K'_m \left( m_e a \right)}{K_m \left( m_e a \right)}, \label{eq:disper}
\end{equation}
where $v_{{\rm A} i,e} = B_0/\sqrt{\mu \rho_{i,e}}$ is the internal/external Alfv\'en speed, and the quantities $m_i$ and $m_e$ are given by
\begin{equation}
m_i^2 = \frac{\left( \omega^2 - k_z^2 \csp^2 \right)  \left( \omega^2 - k_z^2 \vap^2  \right)  }{\left( \csp^2 + \vap^2 \right) \left( \ctp^2 k_z^2 - \omega^2 \right) }, \qquad  m_e^2 = \frac{\left( \omega^2 - k_z^2 \csco^2 \right)  \left( \omega^2 - k_z^2 \vac^2  \right)  }{\left( \csco^2 + \vac^2 \right) \left( \omega^2 - \ctc^2 k_z^2 \right) },
\end{equation}
where $c_{{\rm T} i,e} = c_{s i,e} v_{{\rm A} i,e} / \sqrt{c_{s i,e}^2 + v_{{\rm A} i,e}^2 }$ is the internal/external cusp (or tube) speed.

When an inhomogeneous transitional layer is present in the model, we cannot obtain an analytical expression for the dispersion relation unless some assumptions are made. An elegant method for obtaining an analytical dispersion relation is to combine the jump conditions and the approximation of the thin boundary (TB). The jump conditions were derived by \citet{SGH91} and \citet{goossens95} for the driven problem and by \citet{tirry} for the eigenvalue problem. This method was used, for example, by \citet{sakuraib} for determining the absorption of sound waves in sunspots, \citet{goossens92} for determining surface eigenmodes in incompressible and compressible plasmas, and  \citet{tom} for kink eigenmodes in pressureless coronal loops. This method is valid if the inhomogeneous length-scale within the resonant layer is sufficiently small in comparison with the tube radius. Apart from the Alfv\'en resonance, \citet{SGH91} also provided jump conditions for the slow resonance, which can be applied to the present situation. Hence, the TB formalism allows us to assess the particular contribution of the slow resonance to the kink mode damping.

In our model the magnetic field is straight and constant so that the variations of the local Alfv\'en frequency and the local cusp (or slow) frequency are only due to the variation of density. In that case the jump conditions at the Alfv\'en resonance point, namely $r=r_{\rm A}$, can be written as
\begin{equation}
  \left[ \xi_r \right] = -i \pi \frac{m^2 / r_{\rm A}^2}{\omega_{\rm A}^2 \left| \pd_r \rho_0 \right|_{\rm A}} p_{\rm T} , \quad \left[ p_{\rm T} \right] = 0, \quad {\rm at} \quad r = r_{\rm A}, \label{eq:jumpalfven}
\end{equation}
where $\omega_{\rm A}$ and $\left| \pd_r \rho_0 \right|_{\rm A}$ are the Alfv\'en frequency and the modulus of the radial derivative of the density profile, respectively, both quantities evaluated at $r=r_{\rm A}$. The respective jump conditions at the slow resonance point, namely $r_s$, are
\begin{equation}
  \left[ \xi_r \right] = -i \pi \frac{k_z^2}{\omega_c^2 \left| \pd_r \rho_0 \right|_s} \left( \frac{\cs^2}{\cs^2 + \va^2} \right)^2 p_{\rm T}, \quad \left[ p_{\rm T} \right] = 0, \quad {\rm at} \quad r = r_s, \label{eq:jumpslow}
\end{equation}
where $\omega_c$ and $\left| \pd_r \rho_0 \right|_s$ are the cusp frequency and the modulus of the radial derivative of the density profile, respectively, both quantities evaluated at $r=r_s$, while the factor $\left( \frac{\cs^2}{\cs^2 + \va^2} \right) = \left( \frac{\beta}{\beta + 2/\gamma} \right) \approx 0.034$ is a constant everywhere in the equilibrium, with $\gamma = 5/3$ the adiabatic index. The jump conditions~(\ref{eq:jumpslow}) are independent of the azimuthal wavenumber, $m$, but they depend on the longitudinal wavenumber, $k_z$. So, for $k_z = 0$, jump conditions~(\ref{eq:jumpslow}) become $\left[ \xi_r \right] = \left[ p_{\rm T} \right] = 0$, meaning that there is no slow resonance in such a case.

Next, we use jump conditions~(\ref{eq:jumpalfven}) and (\ref{eq:jumpslow}) to obtain a correction to the dispersion relation~(\ref{eq:disper}) due to both resonances in the TB approximation,
\begin{equation}
 \mathcal{D}_m \left( \omega , k_z \right) = -i \pi \frac{m^2 / r_{\rm A}^2}{\omega_{\rm A}^2} \frac{\rho_i \rho_e}{\left| \pd_r \rho_0 \right|_{\rm A}} -i \pi \frac{k_z^2 }{\omega_c^2} \left( \frac{\cs^2}{\cs^2 + \va^2} \right)^2 \frac{\rho_i \rho_e}{\left| \pd_r \rho_0 \right|_s} . \label{eq:disperalfven}
\end{equation}
The first term on the right-hand side of equation~(\ref{eq:disperalfven}) corresponds to the effect of the Alfv\'en resonance, while the second term is present due to the slow resonance.

\subsubsection{Expressions in the thin tube approximation}

Further analytical progress can be made by combining the thin boundary (TB) and the thin tube (TT) approximations. This was done extensively by \citet{goossens92}. We perform a asymptotic expansion of the Bessel functions present in the dispersion relation~(\ref{eq:disperalfven}) by considering the long-wavelength limit, i.e., $k_z a \ll 1$, and only keep the lowest order, nonzero term of the expansion. Thus, the  dispersion relation~(\ref{eq:disperalfven}) becomes,
\begin{equation}
 \rho_e  \frac{m / a}{\left( \omega^2 - k_z^2 \vap^2 \right)} 
 + \rho_i  \frac{m / a}{\left( \omega^2 - k_z^2 \vac^2 \right)}  + i \pi \rho_i \rho_e \left[ \frac{m^2 / r_{\rm A}^2}{\omega_{\rm A}^2} \frac{1}{\left| \pd_r \rho_0 \right|_{\rm A}} + \frac{k_z^2 }{\omega_c^2} \left( \frac{\cs^2}{\cs^2 + \va^2} \right)^2 \frac{1}{\left| \pd_r \rho_0 \right|_s} \right] = 0. \label{eq:drttap}
\end{equation}

Now, we write the frequency as $\omega = \omega_{\rm R} + i \omega_{\rm I}$. It follows from the resonant conditions that $\omega_{\rm A} = \omega_c = \omega_{\rm R}$. The position of the Alfv\'en, $r_{\rm A}$, and slow,  $r_s$, resonance points can be computed by equating the real part of the kink mode frequency to the local Alfv\'en and slow frequencies, respectively. By this procedure, we obtain
\begin{equation}
 r_{\rm A} = a + \frac{l}{\pi} \arcsin \left[ \frac{\rho_i - \rho_e}{\rho_i + \rho_e} - \frac{2 \vap^2 k_z^2}{\omega_{\rm R}^2} \frac{\rho_i }{\left( \rho_i + \rho_e \right)}  \right], \label{eq:resApoint}
\end{equation}
for the Alfv\'en resonance point, and 
\begin{equation}
 r_s = a + \frac{l}{\pi} \arcsin \left[ \frac{\rho_i - \rho_e}{\rho_i + \rho_e} - \frac{2 \csp^2 k_z^2}{\omega_{\rm R}^2} \frac{\rho_i }{\left( \rho_i + \rho_e \right)}  \right], \label{eq:resSpoint}
\end{equation}
for the slow resonance point. In order to derive equation~(\ref{eq:resSpoint}), we have used the approximation $\omega_c^2 \approx \cs^2 k_z^2$, which is valid for $\cs^2 \ll \va^2$. Note that we need the value of $\omega_{\rm R}$ to determine $r_{\rm A}$  and $r_s$. Expressions for $\left| \pd_r \rho_0 \right|_{\rm A}$ and $\left| \pd_r \rho_0 \right|_s$ are obtained from the density profile (eq.~[\ref{eq:profile}]),
\begin{equation}
 \left| \pd_r \rho_0 \right|_{\rm A}  =\left( \frac{\rho_i - \rho_e}{l} \right) \frac{\pi}{2}\cos \alpha_{\rm A},  \quad \left| \pd_r \rho_0 \right|_s =\left( \frac{\rho_i - \rho_e}{l} \right) \frac{\pi}{2}\cos \alpha_s.\label{eq:proal}
\end{equation}
with $\alpha_{\rm A} = \pi \left( r_{\rm A} - a \right)/l$ and $\alpha_s = \pi \left( r_s- a \right) / l$. We insert these expressions in equation~(\ref{eq:drttap}) and neglect terms with $\omega_{\rm I}^2$, i.e., low-damping situation. Then we obtain an expression for the ratio $\omega_{\rm R}/\omega_{\rm I}$ after some algebraic manipulations. Since the oscillatory period, $P$, and the damping time, $\tau_{\rm D}$, are related to the frequency as follows
\begin{equation}
 P = \frac{2 \pi}{\omega_{\rm R}}, \quad \tau_{\rm D} = \frac{1}{\omega_{\rm I}},
\end{equation}
it is then straight-forward to give an expression for $\tau_{\rm D} / P$,
\begin{equation}
 \frac{\tau_{\rm D}}{P}=\mathcal{F} \frac{1}{(l/a)} \left( \frac{\rho_i + \rho_e}{\rho_i - \rho_e} \right) \left[ \frac{m}{\cos \alpha_{\rm A}}+\frac{\left( k_z a \right)^2}{m}  \left( \frac{\cs^2}{\cs^2 + \va^2} \right)^2 \frac{1}{\cos \alpha_s} \right]^{-1}, \label{eq:ttltbA}
\end{equation}
where $\mathcal{F}$ is a numerical factor that depends on the density profile ($\mathcal{F} = 2/\pi$ in our case). As in previous expressions, the term with $k_z$ corresponds to the contribution of the slow resonance. If this term is dropped and $m=1$ and $\cos \alpha_{\rm A} = 1$, equation~(\ref{eq:ttltbA}) is equivalent to equation~(3) of \citet{arregui}, which only takes the Alfv\'en resonance into account and was first obtained by \citet{goossens92} \citep[see also equivalent expressions in][]{rudermanroberts,goossens02}.

Next, we assume $r_{\rm A} \approx r_s \approx a$ for simplicity, so $\cos \alpha_{\rm A} \approx \cos \alpha_s  \approx 1$. The ratio of the two terms in equation~(\ref{eq:ttltbA}) is
\begin{equation}
 \frac{\left(\tau_{\rm D}\right)_{\rm A}}{\left(\tau_{\rm D}\right)_s} \approx \frac{\left( k_z a \right)^2}{m^2} \left( \frac{\cs^2}{\cs^2 + \va^2} \right)^2, \label{eq:ratiotaus}
\end{equation}
where $\left(\tau_{\rm D}\right)_{\rm A}$ and $\left(\tau_{\rm D}\right)_s$ are the damping times exclusively due to the Alfv\'en and slow resonances, respectively. To make a simple calculation we note that the observed wavelengths of prominence oscillations correspond to $10^{-3} < k_z a < 10^{-1}$. For $m=1$ and $k_z a = 10^{-2}$ we obtain $\left(\tau_{\rm D}\right)_{\rm A} / \left(\tau_{\rm D}\right)_s \approx 10^{-7}$, meaning that the Alfv\'en resonance is much more efficient than the slow resonance for damping the kink mode.  Note that even in the extreme case of a very large $\beta$, i.e., $\cs^2 \to \infty$ and so $\left( \frac{\cs^2}{\cs^2 + \va^2} \right) \to 1$, $\left(\tau_{\rm D}\right)_{\rm A} / \left(\tau_{\rm D}\right)_s \ll 1$ for typical values of $k_z a$. By means of this simple calculation, we can anticipate that the slow resonance will be irrelevant for the kink mode damping.  
\subsection{Numerical computations}
\label{sec:num}
In addition to the analytical approximations, we also numerically solve the full eigenvalue problem by means of the PDE2D code \citep{sewell}. We consider the resistive version of the linear MHD equations. The magnetic Reynolds number, $R_m$, in the corona is considered to be around $10^{12}$, but using this value requires taking an enormous number of grid points in the numerical computations. We therefore use a smaller value of $R_m$, and consequently a smaller number of grid points, but we make sure that the resonant plateau is reached and the damping time becomes independent of the magnetic diffusivity. We use a nonuniform grid with a large density of grid points within the inhomogeneous layer in order to correctly describe the small spatial scales that develop due to both resonances. In the next Section, these numerical results are compared with the analytical TB approximations.

\begin{figure*}[!ht]
\centering
\epsscale{0.325}
\plotone{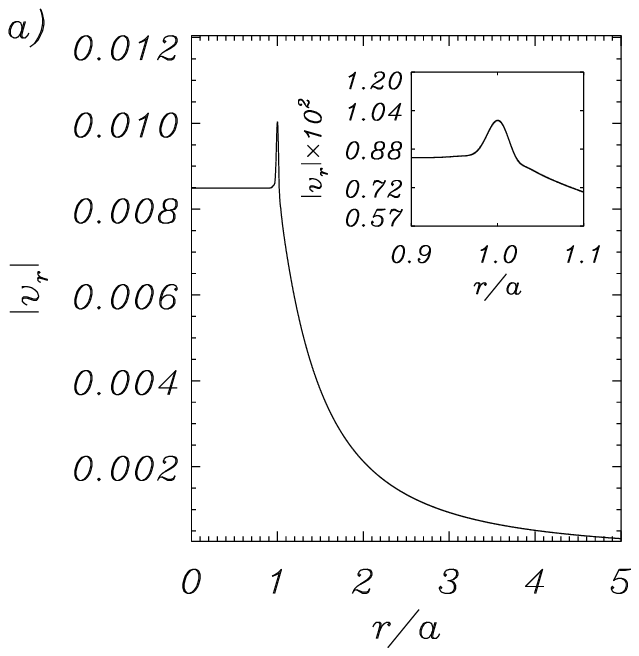}
\plotone{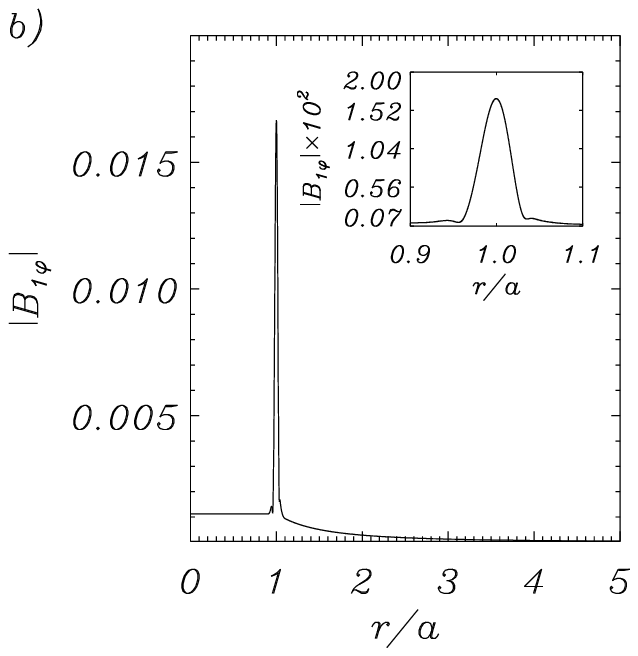}
\plotone{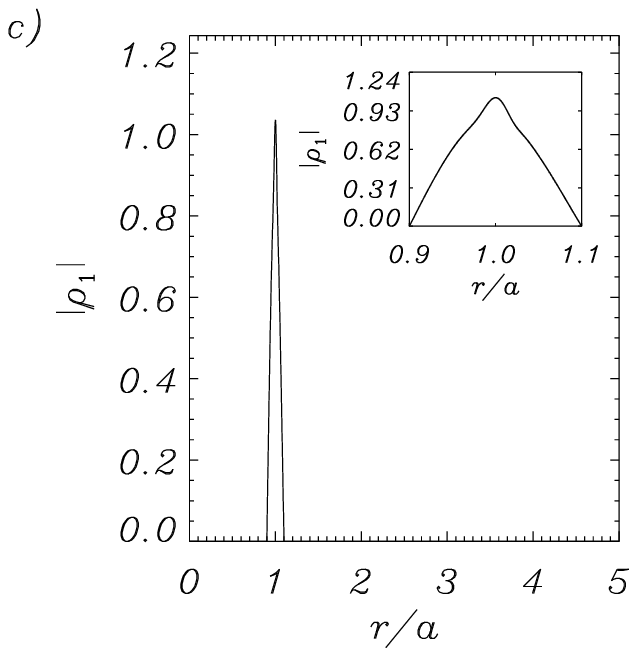}
\plotone{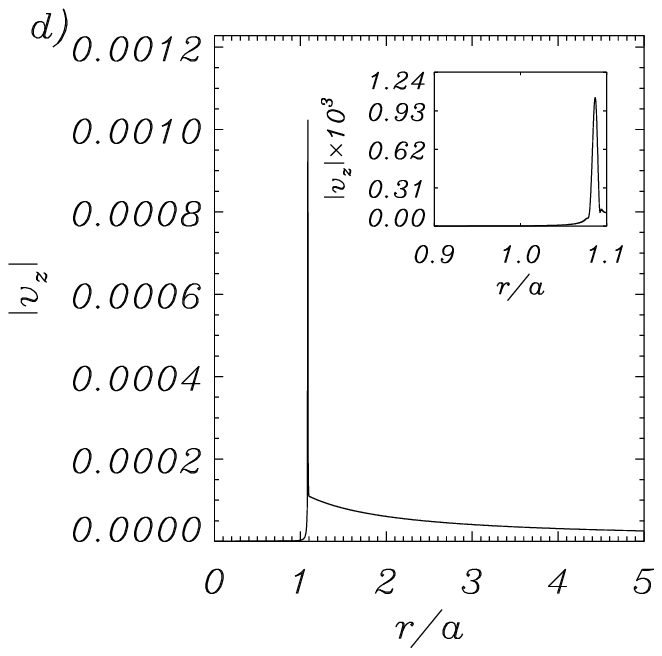}
\plotone{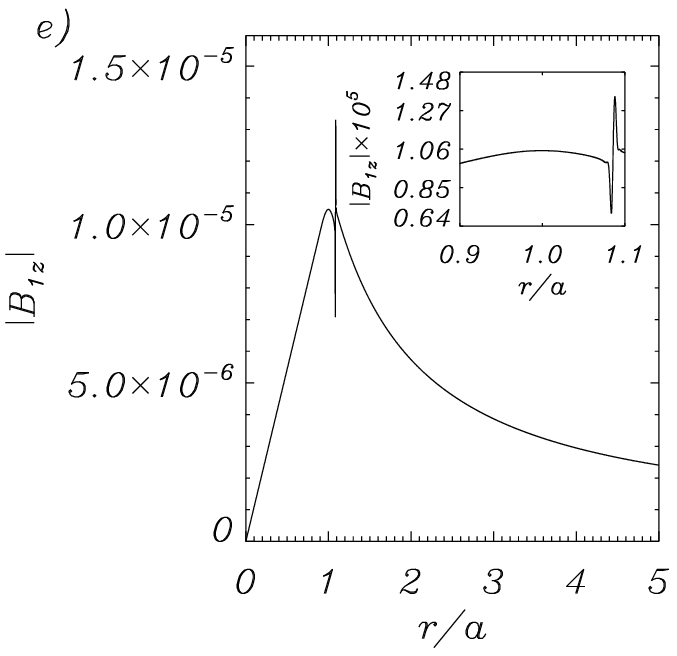}
\plotone{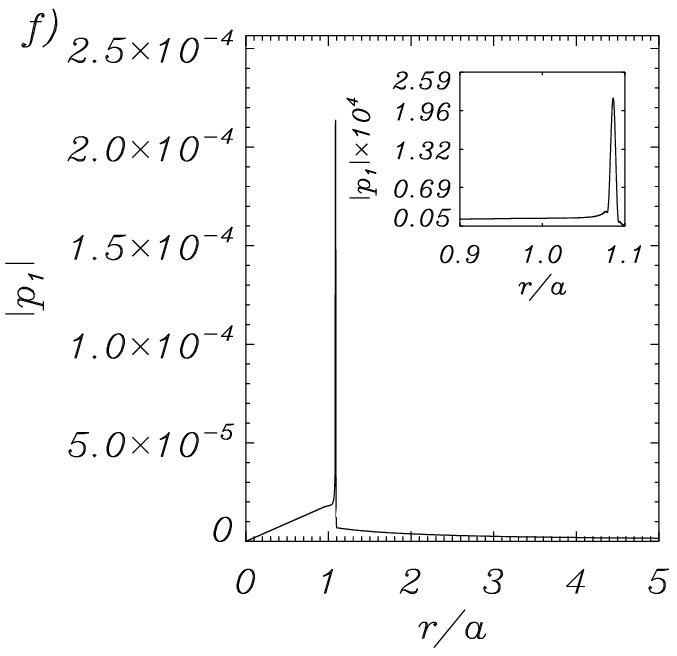}
\caption{Modulus in arbitrary units of kink mode perturbations: $a)$ $v_r$, $b)$ $B_{1 \varphi}$, $c)$ $\rho_1$, $d)$ $v_z$, $e)$ $B_{1 z}$, and $f)$ $p_1$, as functions of $r/a$ for $l/a = 0.2$ and $k_z a = 10^{-2}$. The small panel in each graphic corresponds to an enhancement of the eigenfunction within the transitional layer. \label{fig:eigenfunctions}}
\end{figure*}

\section{Results}
\label{sec:results}

We focus our investigation on the damping of the kink mode ($m=1$). First, we fix the longitudinal wavenumber to $k_z a = 10^{-2}$ and consider $l/a = 0.2$. We numerically compute (see Fig.~\ref{fig:eigenfunctions}) the eigenfunctions: the velocity perturbation ($v_r$, $v_\varphi$, $v_z$), the magnetic field perturbation ($B_{1r}$, $B_{1\varphi}$, $B_{1z}$), the density perturbation ($\rho_1$), and the gas pressure perturbation ($p_1$). The behavior of $v_\varphi$ and $B_{1r}$ is similar to that of $B_{1\varphi}$ and $v_r$, respectively, and hence they are not displayed in Figure~\ref{fig:eigenfunctions}. Because of the resonances the perturbations show large peaks in the transitional layer. The small panels of Figure~\ref{fig:eigenfunctions} display the behavior of the eigenfunctions within the inhomogeneous layer, which allows us to ascertain the position of the peaks in more detail. The peaks of the perturbations $v_r$, $B_{1\varphi}$, and $\rho_1$ are related to the Alfv\'en resonance, while the perturbations $v_z$, $B_{1z}$, and $p_1$ are more affected by the slow resonance and their peaks appear at a different position. Moreover, we see that peaks related to the Alfv\'en resonance are wider than those related to the slow resonance, meaning that the slow resonance produces smaller spatial scales within the resonant layer. Considering the numerical value of $\omega_{\rm R}$, we get $r_{\rm A} / a \approx 1$ and $r_s / a \approx 1.08$ from equations~(\ref{eq:resApoint}) and (\ref{eq:resSpoint}), respectively. Both values are in a good agreement with the position of peaks in Figure~\ref{fig:eigenfunctions}.

\begin{figure}[!ht]
\centering
\epsscale{0.5}
\plotone{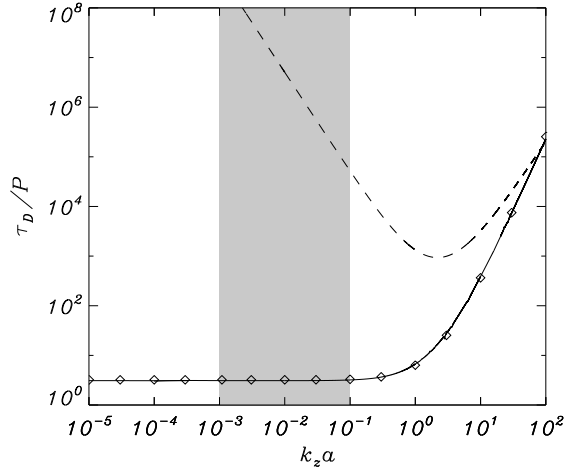}
\caption{Ratio of the damping time to the period, $\tau_{\rm D} / P$, as a function of the dimensionless wavenumber, $k_z a$, corresponding to the kink mode for $l/a = 0.2$. The solid line is the complete numerical solution obtained with the PDE2D code. The symbols and the dashed line are the results of the TB approximation corresponding to the Alfv\'en and slow resonances (the effect of the first and second terms on the right-hand side of equation~[\ref{eq:disperalfven}], respectively). The shaded zone corresponds to the range of typically observed wavelengths in prominence oscillations.  \label{fig:ratio}}
\end{figure}

Next, we plot in Figure~\ref{fig:ratio} the ratio of the damping time to the period, $\tau_{\rm D} / P$, as a function of $k_z a$ corresponding to the kink mode for $l/a = 0.2$. In this Figure, we compare the numerical results with those obtained from the TB approximation. At first sight, we see that the slow resonance (dashed line in Fig.~\ref{fig:ratio}) is much less efficient than the Alfv\'en resonance (symbols) in damping the kink mode. The individual contribution of each resonance in the TB approximation has been determined by solving equation~(\ref{eq:disperalfven}) and only taking the term on the right-hand side related to the desired resonance into account. For the wavenumbers relevant to prominence oscillations, $10^{-3} < k_z a < 10^{-1}$, the value of $\tau_{\rm D} / P$ related to the slow resonance is between 4 and 8 orders of magnitude larger than the ratio obtained by the Alfv\'en resonance. The result of the thin tube limit (eq.~[\ref{eq:ratiotaus}]) agrees with this. On the other hand, the complete numerical solution (solid line) is close to the result for the Alfv\'en resonance. In agreement with \citet{arregui}, we obtain $\tau_{\rm D} / P \approx 3$ in the relevant range of $k_z a$. As was stated by \citet[Fig.~2$d$]{arregui}, the discrepancy between the numerical result and the TB approximation increases with $l/a$, the difference being around 20\% for $l/a = 1$. However, for small, realistic values of $l/a$ this discrepancy is less important and the TB approach is a very good approximation to the numerical result. For short wavelengths ($k_z a \gtrsim 10^0$) the value of $\tau_{\rm D} / P$ increases and the efficiency of the Alfv\'en resonance as a damping mechanism decreases. For $k_z a \approx 10^2$ both the slow and the Alfv\'en resonances produce similar (and inefficient) damping times.

\section{Conclusion}
\label{sec:conclusion}

In this Letter, we have studied the kink mode damping in filament threads due to resonant coupling to slow continuum modes. As far as we know, this is the first time that this phenomenon is studied in the context of solar prominences. By considering the thin boundary approximation, we have found that, contrary to the Alfv\'en resonance, the slow resonance is very inefficient in damping the kink mode for typical prominence-like conditions.  This conclusion also holds for fluting modes ($m \geq 2$). The very small damping due to the slow resonance is comparable to that due to thermal effects studied in previous papers \citep{solerapj}. Therefore, we conclude that the effect of the slow resonance is not relevant to the damping of transverse thread oscillations, which is probably governed by the Alfv\'en resonance.

\acknowledgements{
     RS, RO, and JLB acknowledge the financial support received from the Spanish MICINN and FEDER funds, and the Conselleria d'Economia, Hisenda i Innovaci\'o of the CAIB under Grants No. AYA2006-07637 and PCTIB-2005GC3-03, respectively. RS, RO, and JLB also want to acknowledge the International Space Science Institute teams ``Coronal waves and Oscillations'' and ``Spectroscopy and Imaging of quiescent and eruptive solar prominences from space'' for useful discussions. RS thanks the Conselleria d'Economia, Hisenda i Innovaci\'o for a fellowship. MG acknowledges support from the grant GOA/2009-009. MG is grateful to UIB for financial support during his stay at UIB and to JLB for the hospitality at the Solar Physics Group.}


\begin{thebibliography}{}

%
  \bibitem[Arregui et al.(2008)]{arregui} Arregui, I., Terradas, J. Oliver, R., \& Ballester, J. L. 2008, \apj, 682, L141 
%
%
     \bibitem[Ballester(2006)]{ballester} Ballester, J. L. 2006, Phil. Trans. R. Soc. A, 364, 405 
  \bibitem[Banerjee et al.(2007)]{banerjee} Banerjee, D., Erd\'elyi, R., Oliver R., \& O'Shea, E. 2007, \solphys, 246, 3
%

    \bibitem[Carbonell et al.(2004)]{carbonell} Carbonell, M., Oliver, R., \& Ballester, J. L. 2004, \aap, 415, 739 


\bibitem[\v Cade\v z et al.(1997)]{cadez} \v Cade\v z, V. M., Cs\'ik, \'A., Erd\'elyi, R., \& Goossens, M. 1997, \aap, 326, 1241 


\bibitem[Edwin \& Roberts(1983)]{edwinroberts} Edwin, P. M., \& Roberts, B. 1983, \solphys, 88, 179

\bibitem[Erd\'elyi et al.(2001)]{erdelyi} Erd\'elyi, R., Ballai, I., Goossens, M. 2001, \aap, 368, 662

\bibitem[Goossens et al.(1992)]{goossens92} Goossens, M., Hollweg, J. V., \& Sakurai, T. 1992, \solphys, 138, 233

\bibitem[Goossens et al.(1995)]{goossens95} Goossens, M., Ruderman, M. S., \& Hollweg, J. V. 1995, \solphys, 157, 75

\bibitem[Goossens et al.(2002)]{goossens02} Goossens, M., Andries, J., \& Aschwanden, M. J. 2002, \aap, 394, L39

\bibitem[Goossens et al.(2006)]{goossens} Goossens, M., Andries, J., \& Arregui, I. 2006, Phil. Trans. R. Soc. A, 364, 433

\bibitem[Goossens(2008)]{goossens08} Goossens, M. 2008, in IAU Symp. 247, Waves \& Oscillations in the Solar Atmosphere: Heating and Magneto-Seismology, ed. R. Erd\'elyi \& C. A. Mendoza-Brice\~no (Cambridge: Cambridge Univ. Press), 228



\bibitem[Keppens(1996)]{keppens} Keppens, R. 1996, \apj, 468, 907

    \bibitem[Lin et al.(2003)]{lin2003} Lin, Y., Engvold, O., \& Wiik, J. E. 2003, \solphys, 216, 109
      \bibitem[Lin et al.(2005)]{lin2005} Lin, Y.,  Engvold, O., Rouppe van der Voort, L. H. M., Wiik, J. E., \& Berger, T. E. 2005, \solphys, 226, 239
     \bibitem[Lin et al.(2007)]{lin2007} Lin, Y., Engvold, O., Rouppe van der Voort, L. H. M., \& van Noort, M.  2007, \solphys, 246, 65
%
     \bibitem[Molowny-Horas et al.(1999)]{molowny} Molowny-Horas, R., Wiehr, E., Balthasar, H., Oliver, R., \& Ballester, J. L. 1999, JOSO Annual Report 1998, Astronomical Institute Tatranska Lomnica, 126
%
 \bibitem[Okamoto et al.(2007)]{okamoto} Okamoto, T. J, et al. 2007, Science, 318, 1557 
%

\bibitem[Oliver(2008)]{oliver} Oliver, R. 2008, in IAU Symp. 247, Waves \& Oscillations in the Solar Atmosphere: Heating and Magneto-Seismology, ed. R. Erd\'elyi \& C. A. Mendoza-Brice\~no (Cambridge: Cambridge Univ. Press), 158

     \bibitem[Oliver \& Ballester(2002)]{oliverballester02} Oliver, R., \& Ballester, J. L. 2002, \solphys, 206, 45

\bibitem[Ruderman \& Roberts(2002)]{rudermanroberts} Ruderman, M., \& Roberts, B. 2002, \apj, 577, 475

\bibitem[Sakurai et al.(1991a)]{SGH91} Sakurai, T., Goossens, M, \& Hollweg, J. V. 1991a, \solphys, 133, 227
\bibitem[Sakurai et al.(1991b)]{sakuraib} Sakurai, T., Goossens, M, \& Hollweg, J. V. 1991b, \solphys, 133, 247
 \bibitem[Sewell(2005)]{sewell} Sewell, G. 2005, The Numerical Solution of Ordinary and Partial Differential Equations (Hoboken: Wiley \& Sons)
      \bibitem[Soler et al.(2007)]{soler1} Soler, R., Oliver, R., \& Ballester, J. L. 2007,  \aap, 471, 1023 
%
    \bibitem[Soler et al.(2008)]{solerapj} Soler, R., Oliver, R., \& Ballester, J. L. 2008,  \apj, 684, 725 
%
     \bibitem[Terradas et al.(2002)]{terradasobs} Terradas, J., Molowny-Horas, R., Wiehr, E., Balthasar, H., Oliver, R., \& Ballester, J. L. 2002, \aap, 393, 637

 \bibitem[Terradas et al.(2008)]{hinode} Terradas, J., Arregui, I., Oliver, R., \& Ballester, J. L. 2008, \apj, 678, L153

 \bibitem[Tirry \& Goossens(1996)]{tirry} Tirry, W. J., \& Goossens, M. \apj, 471, 501

 \bibitem[Van Doorsselaere et al.(2004)]{tom} Van Doorsselaere, T., Andries, J., Poedts, S., \& Goossens, M. 2004, \apj, 606, 1223

 \end{thebibliography}
\end{document}